\documentclass{article}
\usepackage{graphicx}
\usepackage{natbib}
\title{X-ray observations of black hole sources}
\author{A R Rao \\
{\it Email: arrao@tifr.res.in a.raghu.rao@gmail.com}\\
	Department of Astronomy and Astrophysics  \\
	Tata Institute of Fundamental Research, Mumbai 400005, India \\
		}

\begin{document}
\maketitle

\abstract{X-ray astronomy is closely related to the study of black hole sources. The discovery that some unseen objects, more massive than any degenerate star, emit huge amounts of X-rays helped accept the concept that back holes are present in X-ray binaries. The detection of copious amounts of highly variable X-rays helped the emergence of the paradigm that all Active Galactic Nuclei harboured a supermassive black hole. Since the bulk of the emission in these sources are in X-rays and X-rays are thought to be originating from regions closest to the black holes, it was expected that X-ray observations would yield significant inputs for our understanding of the physical phenomena happening close to black holes like the disk-jet connection and help measure many important parameters like the mass and spin of the black holes. I will trace the developments in this area for the past several decades and,  noting the relatively limited success, stress the need for more sensitive measurements. I will highlight the recent X-ray polarisation measurements of the Galactic black hole candidate source Cygnus X-1 using the CZT Imager instrument of AstroSat and sketch possible future developments.}

\section{Introduction}
\label{sec:intro}


X-ray astronomy started with the serendipitous discovery of the bright X-ray source Sco X-1 during the legendary rocket flight flown for the ostensible reason of measuring reflected X-rays from the moon \citep{Giacconi1962}. Based on the then prevailing wisdom of stellar structure and evolution and the extrapolation from the measured X-rays from the Sun, bright X-ray sources were not expected in the sky. The detection of Sco X-1 and the subsequent detection of several X-ray sources near the Galactic centre region tickled the imagination of the scientific community and nudged them to accept the existence of certain unusually bright X-ray sources \citep{Morrison1967, Giacconi1968}.
It was, however, the launch of UHURU in 1970 that firmly established our understanding of X-ray binaries. Within a few years of UHURU's launch, it was understood and accepted that accretion onto compact objects is the prime energy source responsible for the copious emission of X-rays in these binary sources. In particular, the enigmatic X-ray source Cygnus X-1 was identified with a normal optical star based on its X-ray and radio positions, and the radial velocity measurements of this optical star indicated the presence of a massive unseen companion  \citep{Bolton1975}. Cygnus X-1 was thought to be an accreting black hole source, though, at that time, it was usual to refer to it as a `black hole candidate' because other possibilities could not be unequivocally eliminated \citep{BlumenthalTucker1974}. 
It, however, took a while to extend these concepts to understand Active Galactic Nuclei (AGN) as accreting supermassive black holes \citep{GurskySchwartz1977, Weedman1977}. The unified model for the AGN emerged in the nineties with the basic tenet that the source of energy is accretion and attempts to understand AGN and black hole X-ray binaries (BH-XRB) as manifestations of the accretion phenomena across black hole masses of several orders of magnitude came to be universally accepted \citep{McHardy2006, FenderBelloni2004}.

 Our understanding of accreting black hole sources is indeed intimately related to the development of X-ray observational capabilities. Over the past six decades, the sensitivity of X-ray detectors has leapfrogged several orders of magnitudes, and the diversity of observations in terms of cadence, energy range, and other features like polarisation is truly breathtaking  \citep{SantangeloReview}. These developments, however, are mostly restricted to lower energy X-rays or soft X-rays below $\sim$10 keV and hence probe only some specific features of the X-ray emission mechanisms in these accreting sources. 
 In this article, I will argue that the skewed development in the sensitivity of X-ray detectors favouring low-energy X-rays has hampered a deeper understanding of the accretion phenomena occurring near black holes.

 In section 2, I will give a historical perspective on the development of X-ray observations and their impact on our understanding of black hole sources. In section 3, I will expand on the idea that the X-ray sensitivity improvements are heavily skewed towards low-energy X-rays.
  Section 4 describes the recent measurements of the hard X-ray polarization in the archetypical BH-XRB Cygnus X-1 using the CZT-Imager instrument onboard the AstroSat satellite. In the last section (section 5), conclusions are given, emphasizing suggestions for further work.

\section{Historical perspectives}
\label{sec:hist} 

X-ray astronomy, the child of the space era, proliferated in tandem with the developments in space technology, culminating in the Great Observatory, {\it Chandra}. A large number of X-ray satellites have been flown  \citep[see][for a comprehensive list]{SantangeloReview}, and here I have selected a few of them to highlight the strategic developments at different eras, and these are listed in Table \ref{tab:missions}.  The table segregates these satellites into different eras (separated by horizontal lines in the table) and also gives certain highlights in our understanding of accreting black hole sources. The mass of the instruments, the nominal effective area, and the energy range are also given in the table to give a sense of the complexity of the instruments.

\begin{table}[!t]
\caption{X-ray astronomy satellite missions$^a$}
\label{tab:missions}       
%
%
\begin{tabular}{p{1.5cm}p{1.4cm}p{0.8cm}p{6.5cm}p{2.6cm}}
\hline
Missions & Year & Mass$^b$ & Instruments$^c$& Results$^d$  \\
\hline
UHURU & 1970-73 & $~~$142 &  840; 1.7-18 &X-ray binaries; Cygnus X-1 \\
\hline
HEAO-A & 1977-79 & 2552 & {\bf A1:}   2000; 0.25-25  {\bf A2:}  3000; 0.15-60  &Timing, spectra.\\
               &.              &            & {\bf A3:}  450; 1-13 $~~~~$ {\bf A4:}  220; 15-10000                           &\\
HEAO-B & 1978-81 & 3130   & {\bf HRI:}  20; 0.15-3  $~${\bf IPC:} 10; 0.4-4 &X-rays from all types (AGN). \\
              &.              &            & {\bf SSS:}  200; 0.5-45 $~~~${\bf FPCS:}  1; 0.4-2.6                          &\\
\hline
RXTE & 1995-12  & 3200 & {\bf PCA:} 6500; 2-60 $~${\bf HEXTE:} 1600; 15-250  & QPOs;  jets.  \\
Chandra & 1999-  & 5000 & {\bf HRC:} 200; 0.06-10  $~~~${\bf ACIS:} 600; 0.2-10   & Deep imaging.\\
XMM & 1999-  & 3700 & {\bf EPIC:} 3325; 0.1-15  $~~~$ & Spectra; grating \\
Suzaku & 2005-15  & 1700 & {\bf XRS:} 600; 0.3-12  {\bf XIS:} 1000; 0.2-12  {\bf HXD:} 300; 10-600 & Wide band spectra. \\
\hline
Swift & 2004-  &  1470 & {\bf XRT:} 135; 0.2-10 $~${\bf BAT:} 5200; 15-150 & Soft \& hard X-ray monitoring. \\
NuSTAR & 2012-  & $~~$350 & 847; 3-78   & Spectra \\
AstroSat & 2015-  & 1513 & {\bf SXT:} 90; 0.3-8 $~${\bf LAXPC:} 6500; 3-80 & Higher energy timing. \\
              &               &         &    {\bf CZTI:} 1000; 20-200    &Polarization. \\
HMXT & 2017-  &  2800 &{\bf LE:} 384; 0.7-15 $~${\bf ME:} 952; 5-30& Timing, spectra. \\
            &               &         &    {\bf HE:} 5100; 20-200    & \\
NICER & 2017- & $~~$372 &1900; 0.2-12 & Timing, spectra.\\
IXPE & 2021-  & $~~$330 & 500; 0.3-10 & Polarization. \\
\hline
\end{tabular}
$^a$Details of only a few important missions are given here.\\
$^b$Instrument mass in kg.\\
$^c$Nominal effective area (in cm$^2$), and energy rage (in keV) are given.\\
$^d$Prominent results pertaining to BH-XRBs for each era are given.
\end{table}

\subsection{The early effort}
\label{subsec:early}

Though there was a flurry of activity in X-ray astronomy using rockets and balloons after the momentous rocket flight of 1962, the short duration of exposure in rocket-based experiments and the blockage due to the residual Earth's atmosphere limiting the energy range of observations in balloon-borne experiments resulted in a pretty poor sensitivity for observations  \citep{Morrison1967, Giacconi1968}. X-ray astronomy came of age only after the launch of the UHURU satellite. 

 UHURU, launched in 1970, though modest by current standards (142 kg), significantly changed our views of the X-ray sky due to the continuous observations possible in a satellite platform. UHURU operated in a scan mode and detected a large number of sources, and identified several of them, including X-ray pulsars, the BH-XRB Cyg X-1, and also identified a large number of extragalactic objects like  AGN, QSOs, and clusters of galaxies. In the seventies, several small-scale experiments like Copernicus (OAO-3), ANS, Ariel-5, Ariel-6, and SAS-3 gave useful X-ray observations \citep{SantangeloReview}. Masses of compact objects were estimated in about six X-ray binaries, and evidence for BH in XRBs was found for only three sources, viz., Cyg X-1, LMC X-3, and A0620-00 \citep{Bahcall1978}.
The very fact that copious X-ray emission, including in hard X-rays, is seen in these accreting objects was sufficient to formulate extensive and intricate accretion theories \citep{Shakura1973}.

 The next significant advance in X-ray observations came with the launch of the HEAO satellites. HEAO-A consisted of the most extensive suite of X-ray instruments, whereas HEAO-B (the Einstein Observatory) used X-ray focusing techniques for the first time, thus improving detection sensitivity by several orders of magnitude. X-ray emission was detected from all classes of objects ranging from stellar coronae to clusters of galaxies. Several innovative niche experiments like EXOSAT in 1983 (enabling long uninterrupted broadband observations), Ginga in 1987 (having an all-sky monitor), Rosat in 1990 (full sky survey at low energies), ASCA in 1993 (the first satellite to use an X-ray CCD), though relatively modest in scale as compared to the HEAO satellites, enriched X-ray astronomy with their unique contributions. 

     Due to the availability of all-sky monitors, many X-ray transients were discovered, and several of them were identified as possible black hole candidates. In the mid-nineties, among the 124 known LMXBs, there were 41  transients and  18 BH LMXBs.   
 Radio outbursts were seen in several black hole sources, but the concept of jet emission from BH-XRBs emerged only in the late nineties \citep{TanakaShibazaki1996}.
QPOs were noticed in BH-XRBs in the EXOSAT era, but `the full pattern, if there is one, has not yet emerged'   \citep{vanderKlis1989}. 
 Though Active Galactic Nuclei (AGN) as accreting supermassive black holes were not thought about in the eighties 
\citep{GurskySchwartz1977, Weedman1977}, the unified model of AGN as accreting supermassive black holes was emerging in the nineties due to the extensive studies in the optical and radio wavelengths.

\subsection{The golden era}
\label{subsec:golden}

 The golden era of X-ray astronomy can be considered to be in the late nineties and early this century. The launch of the Great Observatory {\it Chandra} in 1999 and the equally sophisticated {\it XMM-Newton} in the same year established the maturity of X-ray astronomy. The very high-quality imaging by Chandra and very good spectroscopy (including grating) from both these observatories were nicely complemented by the timing studies by {\it RXTE} (1995-2012) and wide band spectral studies by {\it Suzaku} (2005-2015).

In this era, new phenomena at millisecond time scales were discovered in X-ray binaries.
Though relativistic jets were discovered in the Galactic X-ray binary GRS 1915+105 (discovered by the {\it Granat} satellite) before the launch of {\it RXTE}, the all-sky monitoring ability, quick observation sampling and easy availability of data from {\it RXTE} enabled the discovery of many relativistic jet sources and their detailed understanding 
\citep{MirabelRodriguez1999}, including the fascinating object 
GRS 1915+105 
\citep{FenderBelloni2004}. 
There were 20 BH-XRBs with mass measurements, all monitored by RXTE, and almost 
 half discovered by RXTE \citep{RemillardMcClintock2006}.
Such extensive studies helped connect the accretion phenomena in XRBs and AGNs by discovering fundamental relationships between X-ray and radio intensities and masses of black holes across black hole mass ranging from stellar mass to supermassive. The quick pointing ability of RXTE helped very long-term monitoring of bright AGNs, enabling the discovery of the dependence of 
the break time scale in the power spectral density to the mass and luminosity across mass scales   \citep{McHardy2006}. 
It was established that the `physics of disc-jet coupling in XRBs and AGN are closely linked' \citep{FenderBelloni2004}.

\subsection{The current era}
\label{subsec:current} 

The legacy of X-ray astronomy continued in this century by operating several small-scale topically significant experiments, some of which are listed in Table 1. Swift, though a satellite meant for studying Gamma-ray bursts (GRB),  had the wide field hard X-ray monitor, BAT, providing a continuous record of hard X-ray sources, in particular AGN, and the X-ray Telescope, XRT, due to the swift skewing ability of the satellite could provide snapshot fluxes of a large number of X-ray sources. NuSTAR is the first X-ray satellite with a hard X-ray focusing telescope, thus extending the sensitivity leap due to X-ray focusing to hard X-rays (up to 78 keV). AstroSat and HXMT provided sensitive wide-band measurement capabilities, and IXPE is a dedicated X-ray polarisation measuring satellite.
   
The insight and information we have obtained on the concepts of accretion physics from recent studies can be described as `more of the same thing' and incremental information gathering. The expectation that a detailed study of the spectral and timing properties of BH-XRBs would provide answers to the very fundamental astrophysical questions like measuring the parameters of black holes (mass and spin),  and understanding the intricacies of the  Physics in extreme conditions near these objects did not fully materialise. 

The excellent spectral and timing data at low energies enabled the development of intricate models for the putative accretion disk and the Comptonising plasma. Using these models, efforts were made to measure the spin of the black holes in BH-XRBs, and the spin measurements of many sources have been reported \citep{DraghisSpin2023}. When two methods of spin measurements, spectral fitting and timing (QPO) measurements, were applied to the same source, extremely divergent results were obtained    \citep{Mall2Spin2024}. In Cygnus X-1, it was found that 
the measured BH spin was strongly dependent on how the disk is modelled, mainly because the shape of the high-energy tail is relatively complex \citep{ZdziarskiCygSpin2024}. 

 The number of mass measurements in BH-XRBs did not increase significantly.
Most of the BH mass measurements, of course, came from radial velocity measurements in the optical band.  
There were attempts to measure the mass of the black holes in BH-XRBs from X-ray spectral fitting using physics-based accretion theories like the Two Component Accretion Flow (TCAF) model \citep{Chakraba1995, Chakraba1997}. These methods provided reasonable mass estimates for BH-XRBs \citep{ChakrabaMBH2024} and intermediate-mass black holes  \citep{ChakrabaM822022} based on continuum fitting and also by studying the variation of the QPOs. The spectral models, however, are inconsistent with the models evolved over the years for the accretion disk and the Compton cloud (invoking high reflection), and, perhaps due to this, the mass measurements using TCAF have yet to gain wide acceptance by the community.
Even the latest tool of X-ray polarimetric measurements using IXPE though indicated the presence of a combination of thermal disk, coronal, or reflected emission in BH sources, is yet to help in zooming into the correct accretion disk model for BH sources \citep{CaveroIXPE2024}.

 The current status of our understanding of the accretion phenomenon near black holes can be summarised as the one which can `explain' new observations by tweaking the parameters of several existing `models', but has not yet emerged as a widely accepted paradigm with clear predictive powers explaining possible outcomes in settings with widely different initial conditions (like understanding and explaining the disk-jet connection in widely different classes like TDEs and GRBs).
 
  In the next section, I will highlight the limitations of the sensitivity of  X-ray instruments for studying black holes. 
 
\section{X-ray detection sensitivity}
\label{sec:Sensi}
      
      To understand and evaluate the X-ray sensitivity of X-ray telescope/ detector systems, I have chosen data from four of the best telescope systems for X-ray observations. As an example of a bright source, particularly at higher energies ($>$ 50 keV),  I have taken the data for Crab (nebula and pulsar, together). Near simultaneous data were taken from observations made by NuSTAR (the only operating hard X-ray focussing instrument) and  CZT Imager (CZTI) of AstroSat \citep{Bhalerao2017}, which has a box type Coded Aperture Mask (CAM) arrangement enabling simultaneous background measurement \citep{Mithun2024}. The NuSTAR and CZTI data are fit simultaneously with a power law in the 20 -- 200 keV range. The sensitivity of each instrument, deemed as the errors in the data for the uniform logarithmic binning of 50 data channels per decade of energy, is calculated and rescaled as the square root of time in the units of 10 ks to take care of the fact that  the durations of observation are  different for each instrument. The sensitivity numbers, renormalised to milliCrab units,  are shown in Figure 1. 
 
As an example of a fainter source, I have taken the data for the well-known AGN, IC 4329A. Data were taken from observations made by  XMM-Newton and RXTE (simultaneously in  2003) and NuSTAR (non-simultaneous, observed in 2012). 
Data are fit using a model consisting of a power law and reflection component  \citep{2006ApJ...646..783M}.   As was done earlier, the sensitivities of these three instruments, normalised to the Crab flux and scaled to 10 ks observation, are also shown in Figure 1.
  It can be seen that the sensitivity of NuSTAR for  Crab is inferior compared to that for the  AGN because of the limitations of NuSTAR data storage while observing bright sources.
  
  It can be seen that the huge improvement in sensitivity in X-ray astronomy, brought about by the use of focussing techniques, is confined to low energies (a few $\mu$Crab at one keV). In the intermediate energies (10 - 20 keV), the spectral sensitivities are modest and comparable to each other for focussing instruments and large-area collimated detectors (0.1 to 1 milliCrab).
  The sensitivity drops quite sharply above 20 keV, and beyond 50 keV, the sensitivity, in fact, is quite poor.
  It is doubtful that there will be drastic improvements in the spectral sensitivity of X-ray instruments in the future. The area drops drastically with energy for X-ray mirrors, and above 70 keV, it is very challenging to build them. For the collimated detectors, a high particle-induced background is a serious issue. 
  
  These sensitivity limitations seriously affect our understanding of the accretion phenomena near black holes. For example, for the archetypical black hole source Cygnus X-1, it is possible to look at the broad features of the spectral energy distribution from radio to GeV  \citep{Zdziarski2014}, and conclude that the observed fluxes at different wavelengths are consistent with a preferred jet model, but `the available data do not provide unique constraints on the model free parameters' \citep{Zdziarski2014}. The bulk of the emission from the putative accretion disk is in the 10 keV to  500 keV region (in the hard state of the source): the region, as pointed at earlier,  where the X-ray detection sensitivity is relatively poor.
    
    If we accept that the improvement in the X-ray spectral sensitivity in the crucial hard X-ray region is very unlikely to improve dramatically in the near future, one avenue to exploit would be to explore the accretion phenomena with available tools, including the nascent polarimetry tools, albeit of limited sensitivity.

\begin{figure}[t]
\includegraphics[scale=.35]{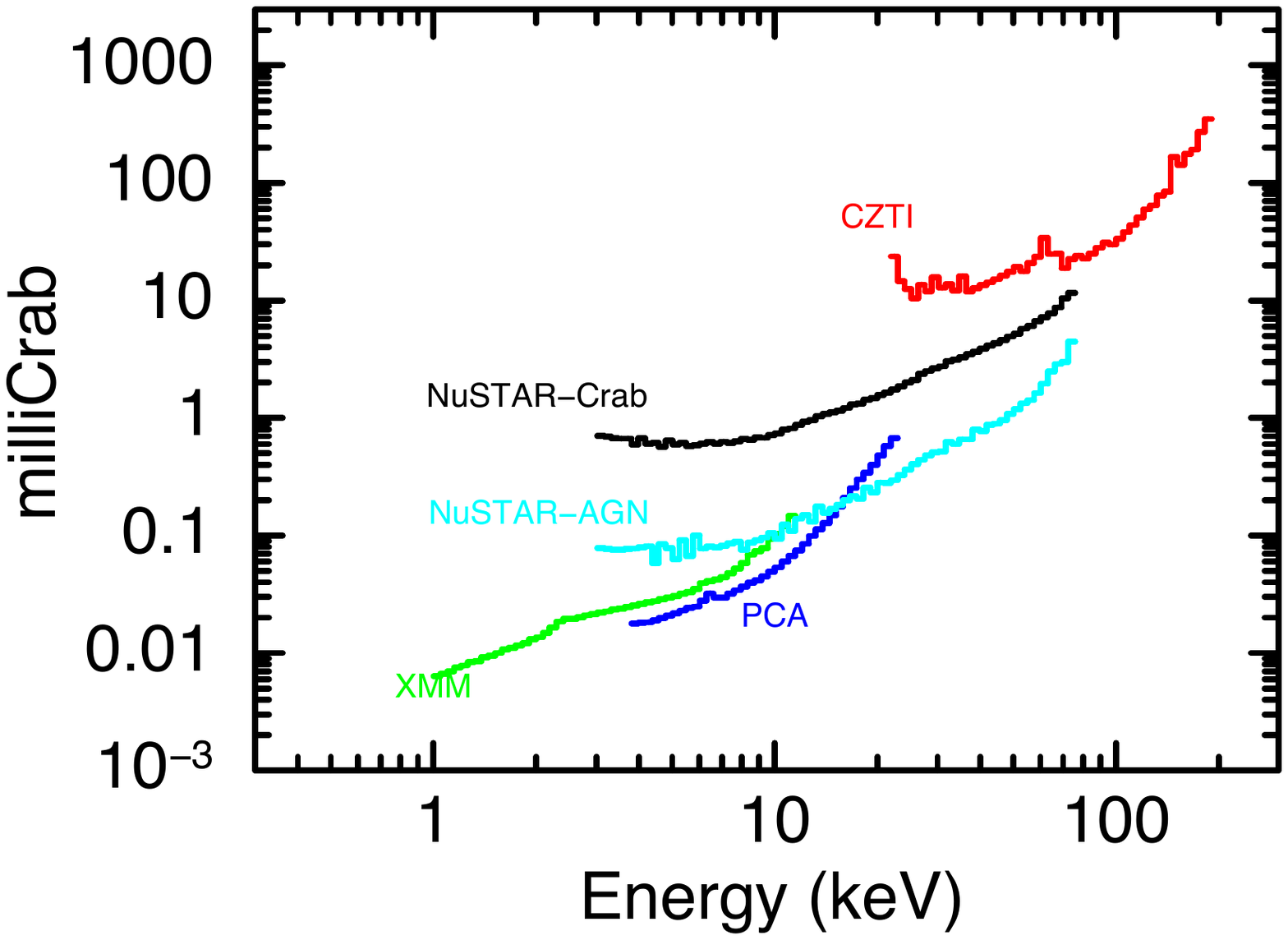}
%
%
\caption{X-ray spectral sensitivity is shown for four instruments. Data for NuSTAR and CZTI of AstroSat are taken from Crab observations, whereas for XMM-Newton-pn, RXTE-PCA and NuSTAR data are taken from observations of a nearby AGN, IC4329A. Sensitivity is taken as the errors in the data for a uniform logarithmic binning of 50 data channels per decade of energy, rescaled as the square root of time in the units of 10 ks, and renormalised to milliCrab units. The sensitivity of NuSTAR for  Crab is inferior to that for the  AGN because of the limitations of NuSTAR data storage while observing bright sources.
}
\label{fig:sensitivity}       
\end{figure}

\subsection{X-ray polarization in Cygnus X-1}       
\label{subsec:CygX1Pol}  

The  IBIS and SPI instruments onboard the INTEGRAL satellite independently measured high polarization for Cygnus X-1 with a polarization degree (PD) of $\sim$ 65\% with the polarization angle (PA)  perpendicular to the jet direction, indicating a strong jet synchrotron component above 400 keV \citep{Laurent2011, Jourdain2012}. 
The balloon-borne hard X-ray polarimeter POGO+, sensitive in the energy range 19 - 181 keV, however,  placed an upper limit of 5.6\% for PD in Cygnus X-1 and also placed an upper limit for polarization from the jet component of around 5\% - 10\%  \citep{Chauvin2018, Chauvin2019}.
The X-ray polarimetric observations of Cygnus X-1 in the hard state using IXPE  (2 - 10 keV) reported a precise measurement of the polarization with a  PD of 4.0\% $\pm$ 0.2\% and an increasing trend of PD  with energy  \citep{Krawczynski2022}.
 The alignment of the X-ray polarization in soft X-rays, however, was with the radio jet (almost perpendicular to the polarisation angle reported by the INTEGRAL data). The precise polarisation data and its energy dependence could be `explained'  with the current ideas of accretion disks. The two variants of the models, viz.,   a hot corona sandwiching the accretion disk or a truncated accretion disk with an inner region of hot corona, could not be distinguished by the available data. 
  
    Recently, the high polarisation at high energies has been independently confirmed by measurements using CZT Imager instruments of AstroSat \citep{Chattopadhyay2024}, which will help us in resolving the tension between the INTEGRAL and the IXPE observations.

\subsection{CZTI measurements of the X-ray polarization in Cygnus X-1}       
\label{subsec:CygX1PolCZTI}  
  
The Cadmium Zinc Telluride Imager (CZTI) instrument of the AstroSat satellite is a collimated hard X-ray detector operating in the 20 - 200 keV range \citep{Bhalerao2017}. The CZTI detectors are pixelated, and the information about any ionization in any of the pixels is recorded, thus making it sensitive to measure Compton scattered events in the detector plane. The scattered events modulate the azimuthal angle distribution if the incident radiation is polarized, and this polarization capability (in the 100 - 380 keV region) was demonstrated in the laboratory  \citep{Chattopadhyay2014, Vadawale2015}. Polarisation measurements using CZTI have been reported for Crab \citep{Vadawale2018} and a large sample of GRBs  \citep{Chattopadhyay2022}.

After the launch of AstroSat in 2015, Cygnus X-1 was observed several times. The initial polarisation results were not very convincing due to the short exposures of each observation and the difficulty of adding data from different exposures. To alleviate these problems, three long ($>$ 200 ks) observations were made, triggered by the transition of the source from soft to hard state (AstroSat Observations ID2992, ID4646, and ID5146). Following the methodology of Lubinski et al. \citep{Lubinski2020}, these three observations were classified as intermediate hard state (HIMS), intermediate soft state (SIMS), and pure hard state (PH), respectively, by examining the variation of the flux with the 20 - 100 keV power-law spectral index (Fig 2a).

\begin{figure}[t]
\includegraphics[scale=.35]{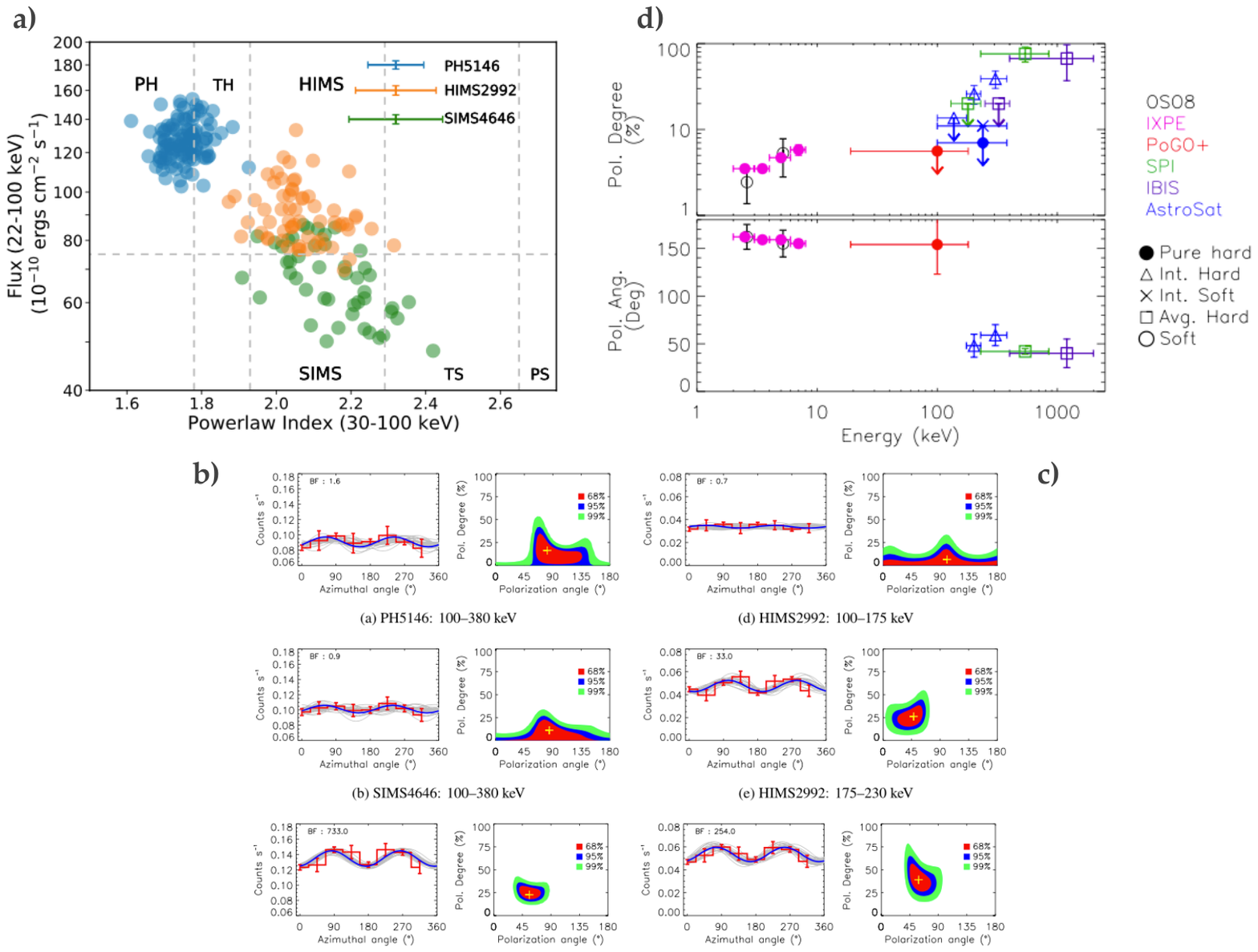}
%
%
\caption{Polarization measurements of Cygnus X-1. a) Three long observations using AstroSat CZTI were segregated into three spectral states of PH, HIMS, and SIMS. b) CZTI polarisation measurements for the three states, the azimuthal distribution of scattered photons shown on the left and the confidence contour levels for the PD and PA, shown on the right. c) Energy resolved polarisation measurements for HIMS. d) A collage of all polarisation measurements of Cygnus X-1, plotted as a function of energy and segregated into spectral states, showing that the CZTI data smoothly joins the low-energy and high-energy measurements \citep[figure adapted from][]{Chattopadhyay2024}.
  }
\label{fig:CygX1}       
\end{figure}

The polarization analysis of these three observations shows exciting results. No evidence for polarization was seen  
in the PH and SIMS states, providing an upper limit of  $\sim$10\%. In the HIMS state, polarized emission was detected with very high ($\sim$6 $\sigma$) significance with a PD of 23$\pm$4\%. An energy resolved analysis hints at resolving the earlier puzzle of low polarisation at low energies and very high polarisation seen at high energies: no polarization was detected below 175 keV, and very high polarisation  (PD $\sim$40\%) is seen at energies above 175 keV, thus smoothly connecting the earlier low and higher energy observations. 

These results are summarised in Fig 2 \citep[adapted from][]{Chattopadhyay2024}. Fig 2a shows the segregation of the three observations into the three states.   The azimuthal scattering angle distributions and the contour plots for  PD and PA are shown for the individual observations (Fig 2b) and for the energy-resolved measurements in the HIMS state (Fig 2c). From all the available measurements, the PD  and PA  of Cygnus X-1 in different spectral states (pure hard, intermediate hard, intermediate soft, averaged hard, and soft states) are shown in Fig 2d.
 We see an apparent increase in PD and a swing in PA at higher energies. 
 
The CZTI measurements confirm the high polarisation at high energies detected by the INTEGRAL satellite. Though this observation sort of resolves the tension between the IXPE and INTEGRAL measurements, understanding of such high polarisation only confined to the higher energies within the premises of our current understanding of the steady state jet emission \citep{Zdziarski2014} still remains an issue. However, the current observations provide several hints at resolving this issue. In the spectral data, there is a signature of a separate spectral component above 100 keV \citep{Chattopadhyay2024}. Further, the high polarization is seen only in the HIMS, a state known for its high and variable radio emission  \citep{Lubinski2020}, indicative of a variable jet component. Hence, the steady-state jet models may not be applicable to the high polarisation measurements seen in this state. Further detailed state-wise observation, simultaneously in broadband (using IXPE, AstroSat, INTEGRAL and Polix) along with radio observations and building good quality state-wise spectral energy distribution along with polarization information will undoubtedly help us in advancing our understanding of the accretion phenomena and the disk jet connection.

\section{Conclusions}
\label{sec:Conc}  

We have given a brief historical overview of the developments in X-ray astronomy, giving particular attention to the advancements in our understanding of the accretion phenomena near black holes. Despite the huge advancement in X-ray detection techniques, the 
behaviour of the innermost region of the accretion disk in black holes has still remained a difficult area of research. It is noted that the technical developments in X-ray astronomy are confined mainly to low energies, thus making observations of limited impact on our understanding of the accretion phenomenon. 

At low energies (2 - 10 keV), the region where the thermal and non-thermal emissions co-exist,  multiple emission processes are possible and, more importantly, since the observational localisation accuracy possible in X-rays (fraction of an arc-second, at best) spans a vast region close to the black hole where multiple hot regions are conceivably possible, any spectrally accurate data can be `understood' in terms of these multiple regions  \citep[if all else fails, put a lamppost !,][]{Lamppost2023}. In hard X-rays, on the other hand, the emission processes are limited and the very fact that these high-energy photons mandate the existence of high-energy electrons with high emissivity demands the emission region very close to the energy generation region, which, in black hole sources, has to be within a few Schwarzschild radii. Thus, any features in hard X-rays necessarily have to come from very close to the black hole with limited emission mechanisms possible.
Though the X-ray sensitivity in the crucial hard X-ray range is limited and unlikely to improve in the near future, X-ray polarisation studies can give important hints about the working of the innermost regions of the accretion disk, including the disk-jet connection.
Comprehensive spectral and polarisation observations in an extremely wide band in X-rays, along with radio observations, will help us to get a firm grip on the accretion phenomenon.

\section*{Acknowledgement}
I thank Gulab Dewangan and N. P. S. Mithun for helping analyse data for the sensitivity calculations. I thank Tanmoy Chattopadhyay for assisting me in making this article's polarisation figure (Fig 2).
 

\bibliography{ref.bib}

\begin{thebibliography}{}
\expandafter\ifx\csname natexlab\endcsname\relax\def\natexlab#1{#1}\fi

\bibitem[{{Bahcall}(1978)}]{Bahcall1978}
{Bahcall}, J.~N. 1978, \araa, 16, 241

\bibitem[{{Bhalerao} {$et~al$.}(2017){Bhalerao}, {Bhattacharya}, {Vibhute},
  {Pawar}, {Rao}, {Hingar}, {Khanna}, {Kutty}, {Malkar}, {Patil}, {Arora},
  {Sinha}, {Priya}, {Samuel}, {Sreekumar}, {Vinod}, {Mithun}, {Vadawale},
  {Vagshette}, {Navalgund}, {Sarma}, {Pandiyan}, {Seetha}, \&
  {Subbarao}}]{Bhalerao2017}
{Bhalerao}, V., {Bhattacharya}, D., {Vibhute}, A., {$et~al$.} 2017, Journal of
  Astrophysics and Astronomy, 38, 31

\bibitem[{{Blumenthal} \& {Tucker}(1974)}]{BlumenthalTucker1974}
{Blumenthal}, G.~R., \& {Tucker}, W.~H. 1974, \araa, 12, 23

\bibitem[{{Bolton}(1975)}]{Bolton1975}
{Bolton}, C.~T. 1975, \apj, 200, 269

\bibitem[{{Cavero} \& {IXPE Collaboration}(2024)}]{CaveroIXPE2024}
{Cavero}, N.~R., \& {IXPE Collaboration}. 2024, in Multifrequency Behaviour of
  High Energy Cosmic Sources XIV, 39

\bibitem[{{Chakrabarti} \& {Titarchuk}(1995)}]{Chakraba1995}
{Chakrabarti}, S., \& {Titarchuk}, L.~G. 1995, \apj, 455, 623

\bibitem[{{Chakrabarti}(1997)}]{Chakraba1997}
{Chakrabarti}, S.~K. 1997, \apj, 484, 313

\bibitem[{{Chattopadhyay} {$et~al$.}(2014){Chattopadhyay}, {Vadawale}, {Rao},
  {Sreekumar}, \& {Bhattacharya}}]{Chattopadhyay2014}
{Chattopadhyay}, T., {Vadawale}, S.~V., {Rao}, A.~R., {Sreekumar}, S., \&
  {Bhattacharya}, D. 2014, Experimental Astronomy, 37, 555

\bibitem[{{Chattopadhyay} {$et~al$.}(2022){Chattopadhyay}, {Gupta}, {Iyyani},
  {Saraogi}, {Sharma}, {Tsvetkova}, {Ratheesh}, {Gupta}, {Mithun}, {Vaishnava},
  {Prasad}, {Aarthy}, {Kumar}, {Rao}, {Vadawale}, {Bhalerao}, {Bhattacharya},
  {Vibhute}, \& {Frederiks}}]{Chattopadhyay2022}
{Chattopadhyay}, T., {Gupta}, S., {Iyyani}, S., {$et~al$.} 2022, \apj, 936, 12

\bibitem[{{Chattopadhyay} {$et~al$.}(2024){Chattopadhyay}, {Kumar}, {Rao},
  {Bhargava}, {Vadawale}, {Ratheesh}, {Dewangan}, {Bhattacharya}, {Mithun}, \&
  {Bhalerao}}]{Chattopadhyay2024}
{Chattopadhyay}, T., {Kumar}, A., {Rao}, A.~R., {$et~al$.} 2024, \apjl, 960, L2

\bibitem[{{Chauvin} {$et~al$.}(2018){Chauvin}, {Flor{\'e}n}, {Friis},
  {Jackson}, {Kamae}, {Kataoka}, {Kawano}, {Kiss}, {Mikhalev}, {Mizuno},
  {Ohashi}, {Stana}, {Tajima}, {Takahashi}, {Uchida}, \&
  {Pearce}}]{Chauvin2018}
{Chauvin}, M., {Flor{\'e}n}, H.~G., {Friis}, M., {$et~al$.} 2018, Nature
  Astronomy, 2, 652

\bibitem[{{Chauvin} {$et~al$.}(2019){Chauvin}, {Flor{\'e}n}, {Jackson},
  {Kamae}, {Kataoka}, {Kiss}, {Mikhalev}, {Mizuno}, {Takahashi}, {Uchida}, \&
  {Pearce}}]{Chauvin2019}
{Chauvin}, M., {Flor{\'e}n}, H.-G., {Jackson}, M., {$et~al$.} 2019, \mnras,
  483, L138

\bibitem[{{Draghis} {$et~al$.}(2023){Draghis}, {Miller}, {Costantini}, {Gallo},
  {Reynolds}, {Tomsick}, \& {Zoghbi}}]{DraghisSpin2023}
{Draghis}, P.~A., {Miller}, J.~M., {Costantini}, E., {$et~al$.} 2023, arXiv
  e-prints, arXiv:2311.16225

\bibitem[{{Fender} \& {Belloni}(2004)}]{FenderBelloni2004}
{Fender}, R., \& {Belloni}, T. 2004, \araa, 42, 317

\bibitem[{{Giacconi} {$et~al$.}(1962){Giacconi}, {Gursky}, {Paolini}, \&
  {Rossi}}]{Giacconi1962}
{Giacconi}, R., {Gursky}, H., {Paolini}, F.~R., \& {Rossi}, B.~B. 1962, \prl,
  9, 439

\bibitem[{{Giacconi} {$et~al$.}(1968){Giacconi}, {Gursky}, \& {van
  Speybroeck}}]{Giacconi1968}
{Giacconi}, R., {Gursky}, H., \& {van Speybroeck}, L.~P. 1968, \araa, 6, 373

\bibitem[{{Gursky} \& {Schwartz}(1977)}]{GurskySchwartz1977}
{Gursky}, H., \& {Schwartz}, D.~A. 1977, \araa, 15, 541

\bibitem[{{Jourdain} {$et~al$.}(2012){Jourdain}, {Roques}, {Chauvin}, \&
  {Clark}}]{Jourdain2012}
{Jourdain}, E., {Roques}, J.~P., {Chauvin}, M., \& {Clark}, D.~J. 2012, \apj,
  761, 27

\bibitem[{{Krawczynski} {$et~al$.}(2022){Krawczynski}, {Muleri},
  {Dov{\v{c}}iak}, {Veledina}, {Rodriguez Cavero}, {Svoboda}, {Ingram}, {Matt},
  {Garcia}, {Loktev}, {Negro}, {Poutanen}, {Kitaguchi}, {Podgorn{\'y}},
  {Rankin}, {Zhang}, {Berdyugin}, {Berdyugina}, {Bianchi}, {Blinov},
  {Capitanio}, {Di Lalla}, {Draghis}, {Fabiani}, {Kagitani}, {Kravtsov},
  {Kiehlmann}, {Latronico}, {Lutovinov}, {Mandarakas}, {Marin}, {Marinucci},
  {Miller}, {Mizuno}, {Molkov}, {Omodei}, {Petrucci}, {Ratheesh}, {Sakanoi},
  {Semena}, {Skalidis}, {Soffitta}, {Tennant}, {Thalhammer}, {Tombesi},
  {Weisskopf}, {Wilms}, {Zhang}, {Agudo}, {Antonelli}, {Bachetti}, {Baldini},
  {Baumgartner}, {Bellazzini}, {Bongiorno}, {Bonino}, {Brez}, {Bucciantini},
  {Castellano}, {Cavazzuti}, {Ciprini}, {Costa}, {De Rosa}, {Del Monte}, {Di
  Gesu}, {Di Marco}, {Donnarumma}, {Doroshenko}, {Ehlert}, {Enoto},
  {Evangelista}, {Ferrazzoli}, {Gunji}, {Hayashida}, {Heyl}, {Iwakiri},
  {Jorstad}, {Karas}, {Kolodziejczak}, {La Monaca}, {Liodakis}, {Maldera},
  {Manfreda}, {Marscher}, {Marshall}, {Mitsuishi}, {Ng},
  {O{\textquoteright}Dell}, {Oppedisano}, {Papitto}, {Pavlov}, {Peirson},
  {Perri}, {Pesce-Rollins}, {Pilia}, {Possenti}, {Puccetti}, {Ramsey},
  {Romani}, {Sgr{\`o}}, {Slane}, {Spandre}, {Tamagawa}, {Tavecchio}, {Taverna},
  {Tawara}, {Thomas}, {Trois}, {Tsygankov}, {Turolla}, {Vink}, {Wu}, {Xie}, \&
  {Zane}}]{Krawczynski2022}
{Krawczynski}, H., {Muleri}, F., {Dov{\v{c}}iak}, M., {$et~al$.} 2022, Science,
  378, 650

\bibitem[{{Laurent} {$et~al$.}(2011){Laurent}, {Rodriguez}, {Wilms}, {Cadolle
  Bel}, {Pottschmidt}, \& {Grinberg}}]{Laurent2011}
{Laurent}, P., {Rodriguez}, J., {Wilms}, J., {$et~al$.} 2011, Science, 332, 438

\bibitem[{{Lubi{\'n}ski} {$et~al$.}(2020){Lubi{\'n}ski}, {Filothodoros},
  {Zdziarski}, \& {Pooley}}]{Lubinski2020}
{Lubi{\'n}ski}, P., {Filothodoros}, A., {Zdziarski}, A.~A., \& {Pooley}, G.
  2020, \apj, 896, 101

\bibitem[{{Mall} {$et~al$.}(2024){Mall}, {Liu}, {Bambi}, {Steiner}, \&
  {Garc{\'\i}a}}]{Mall2Spin2024}
{Mall}, G., {Liu}, H., {Bambi}, C., {Steiner}, J.~F., \& {Garc{\'\i}a}, J.~A.
  2024, \mnras, 527, 12053

\bibitem[{{Markowitz} {$et~al$.}(2006){Markowitz}, {Reeves}, \&
  {Braito}}]{2006ApJ...646..783M}
{Markowitz}, A., {Reeves}, J.~N., \& {Braito}, V. 2006, \apj, 646, 783

\bibitem[{{McHardy} {$et~al$.}(2006){McHardy}, {Koerding}, {Knigge}, {Uttley},
  \& {Fender}}]{McHardy2006}
{McHardy}, I.~M., {Koerding}, E., {Knigge}, C., {Uttley}, P., \& {Fender},
  R.~P. 2006, \nat, 444, 730

\bibitem[{{Mirabel} \& {Rodr{\'\i}guez}(1999)}]{MirabelRodriguez1999}
{Mirabel}, I.~F., \& {Rodr{\'\i}guez}, L.~F. 1999, \araa, 37, 409

\bibitem[{{Mithun}(2024)}]{Mithun2024}
{Mithun}, N.~P.~S. 2024, Private communications, 1

\bibitem[{{Mondal} {$et~al$.}(2022){Mondal}, {Palit}, \&
  {Chakrabarti}}]{ChakrabaM822022}
{Mondal}, S., {Palit}, B., \& {Chakrabarti}, S.~K. 2022, Journal of
  Astrophysics and Astronomy, 43, 90

\bibitem[{{Morrison}(1967)}]{Morrison1967}
{Morrison}, P. 1967, \araa, 5, 325

\bibitem[{{Nath} {$et~al$.}(2024){Nath}, {Debnath}, {Chatterjee}, {Bhowmick},
  {Chang}, \& {Chakrabarti}}]{ChakrabaMBH2024}
{Nath}, S.~K., {Debnath}, D., {Chatterjee}, K., {$et~al$.} 2024, \apj, 960, 5

\bibitem[{{Podgorn{\'y}} {$et~al$.}(2023){Podgorn{\'y}}, {Dov{\v{c}}iak},
  {Goosmann}, {Marin}, {Matt}, {R{\'o}{\.z}a{\'n}ska}, \&
  {Karas}}]{Lamppost2023}
{Podgorn{\'y}}, J., {Dov{\v{c}}iak}, M., {Goosmann}, R., {$et~al$.} 2023,
  \mnras, 524, 3853

\bibitem[{{Remillard} \& {McClintock}(2006)}]{RemillardMcClintock2006}
{Remillard}, R.~A., \& {McClintock}, J.~E. 2006, \araa, 44, 49

\bibitem[{{Santangelo} {$et~al$.}(2023){Santangelo}, {Madonia}, \&
  {Piraino}}]{SantangeloReview}
{Santangelo}, A., {Madonia}, R., \& {Piraino}, S. 2023, arXiv e-prints,
  arXiv:2307.06652

\bibitem[{{Shakura} \& {Sunyaev}(1973)}]{Shakura1973}
{Shakura}, N.~I., \& {Sunyaev}, R.~A. 1973, \aap, 24, 337

\bibitem[{{Tanaka} \& {Shibazaki}(1996)}]{TanakaShibazaki1996}
{Tanaka}, Y., \& {Shibazaki}, N. 1996, \araa, 34, 607

\bibitem[{{Vadawale} {$et~al$.}(2015){Vadawale}, {Chattopadhyay}, {Rao},
  {Bhattacharya}, {Bhalerao}, {Vagshette}, {Pawar}, \&
  {Sreekumar}}]{Vadawale2015}
{Vadawale}, S.~V., {Chattopadhyay}, T., {Rao}, A.~R., {$et~al$.} 2015, \aap,
  578, A73

\bibitem[{{Vadawale} {$et~al$.}(2018){Vadawale}, {Chattopadhyay}, {Mithun},
  {Rao}, {Bhattacharya}, {Vibhute}, {Bhalerao}, {Dewangan}, {Misra}, {Paul},
  {Basu}, {Joshi}, {Sreekumar}, {Samuel}, {Priya}, {Vinod}, \&
  {Seetha}}]{Vadawale2018}
{Vadawale}, S.~V., {Chattopadhyay}, T., {Mithun}, N.~P.~S., {$et~al$.} 2018,
  Nature Astronomy, 2, 50

\bibitem[{{van der Klis}(1989)}]{vanderKlis1989}
{van der Klis}, M. 1989, \araa, 27, 517

\bibitem[{{Weedman}(1977)}]{Weedman1977}
{Weedman}, D.~W. 1977, \araa, 15, 69

\bibitem[{{Zdziarski} {$et~al$.}(2014){Zdziarski}, {Pjanka}, {Sikora}, \&
  {Stawarz}}]{Zdziarski2014}
{Zdziarski}, A.~A., {Pjanka}, P., {Sikora}, M., \& {Stawarz}, {\L}. 2014,
  \mnras, 442, 3243

\bibitem[{{Zdziarski} {$et~al$.}(2024){Zdziarski}, {Chand}, {Banerjee},
  {Szanecki}, {Janiuk}, {Lubinski}, {Niedzwiecki}, {Dewangan}, \&
  {Misra}}]{ZdziarskiCygSpin2024}
{Zdziarski}, A.~A., {Chand}, S., {Banerjee}, S., {$et~al$.} 2024, arXiv
  e-prints, arXiv:2402.12325

\end{thebibliography}
\end{document}